# Partial annotations for the segmentation of large structures with low annotation cost

Bella Specktor Fadida[1], Daphna Link Sourani[2], Liat Ben Sira[3,4], Elka Miller[5], Dafna Ben Bashat[2,3], Leo Joskowicz[1]

[1] School of Computer Science and Engineering, The Hebrew University of Jerusalem, Israel
[2] Sagol Brain Institute, Tel Aviv Sourasky Medical Center, Israel
[3] Sackler Faculty of Medicine & Sagol School of Neuroscience, Tel Aviv University, Israel
[4] Division of Pediatric Radiology, Tel Aviv Sourasky Medical Center, Tel Aviv-Yafo, Israel
[5] Medical Imaging, Children's Hospital of Eastern Ontario, University of Ottawa, Canada

Emails: `bella.specktor@cs.huji.ac.il,josko@cs.huji.ac.il`

**Abstract.** Deep learning methods have been shown to be effective for the automatic segmentation of structures and pathologies in medical imaging. However, they require large annotated datasets, whose manual segmentation is a tedious and time-consuming task, especially for large structures. We present a new method of partial annotations of MR images that uses a small set of consecutive annotated slices from each scan with an annotation effort that is equal to that of only few annotated cases. The training with partial annotations is performed by using only annotated blocks, incorporating information about slices outside the structure of interest and modifying a batch loss function to consider only the annotated slices. To facilitate training in a low data regime, we use a two-step optimization process. We tested the method with the popular soft Dice loss for the fetal body segmentation task in two MRI sequences, TRUFI and FIESTA, and compared full annotation regime to partial annotations with a similar annotation effort. For TRUFI data, the use of partial annotations yielded slightly better performance on average compared to full annotations with an increase in Dice score from 0.936 to 0.942, and a substantial decrease in Standard Deviations (STD) of Dice score by 22% and Average Symmetric Surface Distance (ASSD) by 15%. For the FIESTA sequence, partial annotations also yielded a decrease in STD of the Dice score and ASSD metrics by 27.5% and 33% respectively for in-distribution data, and a substantial improvement also in average performance on out-of-distribution data, increasing Dice score from 0.84 to 0.9 and decreasing ASSD from 7.46 to 4.01 mm. The two-step optimization process was helpful for partial annotations for both in-distribution and out-of-distribution data. The partial annotations method with the two-step optimizer is therefore recommended to improve segmentation performance under low data regime.

# 1  Introduction

Fetal MRI has the potential to complement US imaging and improve fetal development assessment by providing more accurate volumetric information about the fetal structures [1,2]. However, volumetric measurements require manual delineation, also called segmentation, of the fetal structures, which is time consuming, annotator-dependent and error-prone.

In this paper, we focus on the task of fetal body segmentation in MRI scans. Several automatic segmentation methods were proposed for this task. In an early work, Zhang et al. [3] proposed a graph-based segmentation method. More recently, automatic segmentation methods for fetal MRI are based on deep neural networks. Dudovitch et al. [4] describes a fetal body segmentation network that reached high performance with only nine annotated examples. However, the method was tested only on data with similar resolutions and similar gestational ages for the FIESTA sequence. Lo et al. [5] proposed a 2D deep learning framework with cross attention squeeze and excitation network with 60 training scans for fetal body segmentation in SSFP sequences.

While effective, robust deep learning methods usually require a large, high-quality dataset of expert-validated annotations, which is very difficult and expensive to obtain. The annotation process is especially time consuming for structures with large volumes, as they require the delineation of many slices. Therefore, in many cases, the annotation process is performed iteratively, when first initial segmentation is obtained with few annotated datasets, and subsequently manual segmentations are obtained by correcting network results. However, the initial segmentation network trained on few datasets is usually not robust and might fail for cases that are very different from the training set.

To address the high cost associated with annotating structures with large volumes, one approach is to use sparse annotations, where only a fraction of the slices or pixels are annotated [6]. Çiçek et al. [7] describes a 3D network to generate a dense volumetric segmentation from sparse annotations, in which uniformly sampled slices were selected for manual annotation. Goetz et al. [8] selectively annotated unambiguous regions and employed domain adaptation techniques to correct the differences between the training and test data distributions caused by sampling selection errors. Bai et al. [9] proposed a method that starts by propagating the label map of a specific time frame to the entire longitudinal series based on the motion estimation, and then combines FCN with a Recurrent Neural Network (RNN) for longitudinal segmentation. Lejeune et al. [10] introduced a semi-supervised framework for video and volume segmentation that iteratively refined the pixel-wise segmentation within an object of interest. However, these methods impose restrictions on the way the partial annotations are sampled and selected that may be inconvenient for the annotator and still require significant effort.

Wang et al. [11] proposed using incomplete annotations in a user-friendly manner of either a set of consecutive slices or a set of typical separate slices. They used a combined cross-entropy loss with boundary loss and performed labels completion based on the network output uncertainty that was incorporated in the loss function. They showed that their method with 30% of annotated slices was close to the performance using full annotations. However, the authors did not compare segmentation results using full versus partial annotations with the same annotation effort. Also, a question remains if user-friendly partial annotations can be leveraged in the context of the Dice loss as well, a widely used loss function that is robust to class imbalance [12].

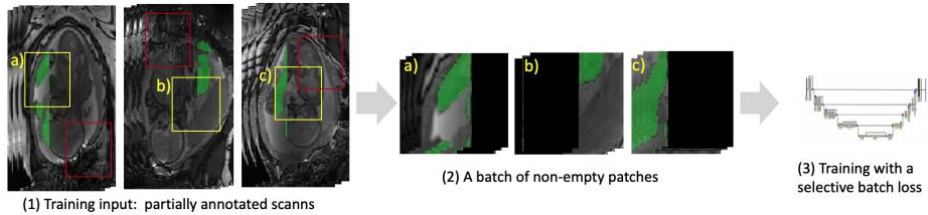

**Fig. 1.** Training flow with partial annotations. 1) Non-empty blocks are picked from the partially annotated scans (sagittal view, example of relevant blocks is shown in yellow). 2) A batch of non-empty blocks is used as input along with information about non-empty slices. The black areas of the blocks correspond to unselected voxels (voxels that are not used by the loss function). 3) The network is trained with a selective loss that uses only the pixels in annotated slices.

Training with limited data usually makes the training optimization more difficult. Therefore, to facilitate optimization, we seek a scheme that will help in avoiding convergence to a poor local minimum. Smith [13] proposed the usage of a cyclic learning rate to remove the need for finding the best values and schedule for the global learning rates. Loshchilov et al [14] showed the effectiveness of using warm learning rate restarts to deal with ill-conditioned functions. They used a simple learning rate restart scheme after a predefined number of epochs.

In this paper, we explore the effectiveness of using partial annotations under low data regime with the Soft Dice loss function. We also explore the usefulness of a warm restarts optimization scheme in combination with fine-tuning to deal with the optimization difficulties under low data regime.

## 2   Method

Our segmentation method with small annotation cost consists of two main steps: 1) manual partial delineations, where the user partially annotates scans with the guidance of the algorithm; 2) training with partial annotations, where a 3D segmentation network is trained with blocks of the partially annotated data using a selective loss function.

The manual partial delineations step is performed as follows. First, the uppermost and lowermost slices of the organ are manually selected by the annotator, which is a quick and easy task. Then, the algorithm randomly chooses a slice within the structure of interest. Finally, the slices to annotate around this slice are selected. The number of slices is determined by the chosen annotation percentage. The annotation percentage is taken from the slices that include the structure of interest, i.e., non-empty segmentation slices. The slices to annotate are chosen consecutively to reduce annotation time, as often the annotations depend on the 3D structure of the organ seen by scrolling and viewing nearby slices during the annotation.

The training with partial annotations is performed as follows. Only the non-empty blocks of the partially annotated data are used for training, as some of the blocks may not include annotations at all. To enrich the annotated data, we also use the border slices information in the loss function and treat the slices outside the structure of interest as annotated slices. We add as input to the network a binary mask specifying the locations of the annotated slices during training. The network is trained with a selective loss function that takes into account only the annotated slices. Also, we use a relatively large batch size of 8 to include enough information during each optimization iteration. Fig. 1 shows the training flow.

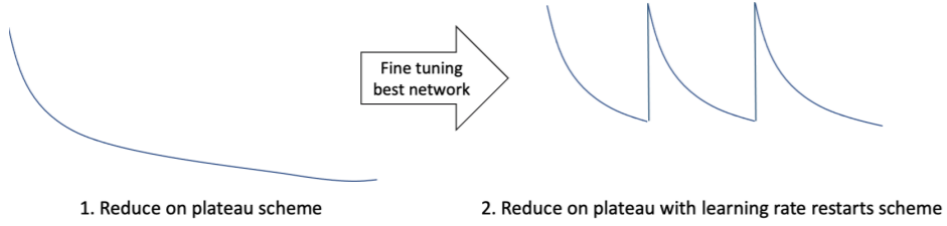

**Fig. 2.** Illustration of the two-step optimization process with the proposed learning rate regimes (graphs of learning rate as a function of epoch number).

### 2.1 Selective Dice Loss

To train a network with partially annotated data, we modify the loss function to use only the annotated slices information. We illustrate the use of a selective loss for the commonly used Soft Dice loss. A batch loss is used, meaning that the calculation is performed on the 4-dimentional batch directly instead of averaging the losses of single data blocks.

Let the number of image patches be $I$ and let the image patch consist of $C$ pixels. The number of voxels in a minibatch is therefore given by $I \times C = N$. Let $t_i$ be a voxel at location $i$ in the minibatch for the ground truth delineation $t_i \in T$ and $r_i$ be a voxel at the location $i$ in the minibatch for the network result $r_i \in R$.

The Batch Dice loss [15] is defined as:

$$Batch\ Dice\ Loss\ (L_{CD}) = -\frac{2\sum_N t_i r_i}{\sum_N t_i + \sum_N r_i} \quad (1)$$

Since we have partial annotations, we will use only the annotated slices locations in the loss calculation. Let $T' \subset T$ and $R' \subset R$ be the ground truth in the annotated slices and the network result in the annotated slices, with minibatch voxels $t_i' \in T'$ and $r_i' \in R'$ respectively. The number of voxels that we consider in the minibatch is now $N' < N$, corresponding only to the annotated slices. The batch dice loss for partial annotations is defined as:

$$Selective\ Batch\ Dice\ Loss\ (L_{CD}) = -\frac{2\sum_{N'} t_i' r_i'}{\sum_{N'} t_i' + \sum_{N'} r_i'} \quad (2)$$

### 2.2 Optimization

To facilitate the optimization process under small data regime, we perform the training in two steps. First, a network is trained with reduction of learning rate on plateau. Then, we use the weights of the network with best results on the validation set to continue training. Similarly to the first phase, the training in the second phase is performed with reduction in plateau, but this time with learning rate restarts every predefined number of epochs (Fig. 2).

## 3   Experimental Results

To evaluate our method, we retrospectively collected fetal MRI scans with the FIESTA and TRUFI sequences and conducted two studies.

**Datasets and annotations**: We collected fetal body MRI scans of patients acquired with the true fast imaging with steady-state free precession (TRUFI) and the fast imaging employing steady-state acquisition (FIESTA) sequences from the Sourasky Medical Center (Tel Aviv, Israel) with gestational ages (GA) 28-39 weeks and fetal body MRI scans acquired with the FIESTA sequence from Children's Hospital of Eastern Ontario (CHEO), Canada with GA between 19-37 weeks. Table 1 shows detailed description of the data.

**Table 1.** Datasets description.

| MRI sequence | ID/OOD | Clinical Site | Scanners | Resolution (mm$^3$) | Pixels/slice | # Slices | GA | # |
|---|---|---|---|---|---|---|---|---|
| **TRUFI** | ID | Sourasky Medical Center | Siemens Skyra 3T, Prisma 3T, Aera 1.5T | 0.6-1.34×0.6-1.34 ×2-4.8 | 320-512 ×320-512 | 50-120 | 28-39 | 101 |
| **FIESTA** | ID | Sourasky Medical Center | GE MR450 1.5T | 1.48-1.87×1.48-1.8 ×2-5 | 256×256 | 50-100 | 28-39 | 104 |
| | OOD | Children's Hospital | Mostly GE Signa HDxt 1.5T; Signa 1.5T, SIEMENS Skyra 3T | 0.55-1.4×0.55-1.4 ×3.1-7.5 | 256×256 512×512 | 19-55 | 19-37, mostly 19-24 | 33 |

Ground truth segmentations were created as follows. First, 36 FIESTA cases were annotated from scratch. Then, 68 ID and 33 OOD cases were manually corrected from network results. For the TRUFI data all cases were created by correcting network results: first, a FIESTA network was used to perform initial segmentation and afterwards a TRUFI network was trained for improved initial segmentation. Both the annotations and the corrections were performed by a clinical trainee. All segmentations were validated by a clinical expert.

**Studies:** We conducted two studies that compare partial annotations to full annotations with the same number of slices. Study 1 evaluates the partial annotations method for the TRUFI body dataset and performs ablation for the two-step optimization process and the usage of slices outside of the fetal body structure. Study 2 evaluates the partial annotations method for the FIESTA body dataset for both ID and OOD data.

For both studies, we compared training with 6 fully annotated cases to 30 partially annotated cases with annotation of 20% of the slices. The selection of cases and the location for partial annotations was random for all experiments. Because of the high variability in segmentation quality for the low-data regime, we performed all the experiments with four different randomizations and averaged between them. The segmentation quality is evaluated with the Dice, Hausdorff and 2D ASSD (slice Average Symmetric Surface Difference) metrics.

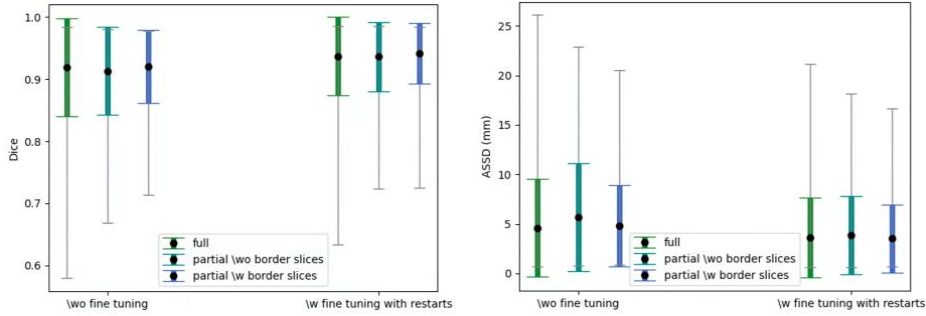

**Fig. 3.** Fetal body segmentation results for the FIESTA sequence. Training with full annotations (full) is compared to training with partial annotations with (\w) and without (\wo) border slices. The colored bars show the STD of the metric and the grey bars show the range of the metric (minimum and maximum).

A network architecture similar to Dudovitch et al. [4] was utilized with a patch size of 128×128×48 to capture a large field of view. A large batch size of 8 was used in all experiments to allow for significant updates in each iteration for the partial annotations regime. Since the TRUFI sequence had a higher resolution compared to FIESTA, the scans were downscaled by ×0.5 in the in-plane axes to have a large field of view [16]. The segmentation results were refined by standard post-processing techniques of holes filling and extraction of the main connected component.

Both partially annotated and fully annotated networks were trained in a two-step process. First, the network was trained with a decreasing learning rate, with an initial learning rate of 0.0005. The network that yielded the best validation result was selected, and this network was then fine-tunned on the same data. For fine-tuning, we again used a decreasing learning rate scheme with an initial learning rate of 0.0005, but this time we performed learning rate restarts every 60 epochs.

*Study 1: partial annotations for TRUFI sequence and ablation*

The method was evaluated on 30/13/58 training/validation/test split for partially annotated cases with 20% of annotated slices and 6/13/58 for fully annotated cases. The 6 fully annotated training cases were randomly chosen out of the 30 partially annotated training cases. Ablation experiments were performed to evaluate the effectiveness of the two-step optimization scheme and the usage of slices outside the body structure.

Six scenarios were tested: 1) full annotations without fine tuning; 2) partial annotations without fine tuning and without borders information; 3) partial annotations without fine-tuning and with borders information; 4) full annotations with fine tuning; 5) partial annotations with fine tuning but without borders information; 6) partial annotations with fine tuning and borders information.

Fig. 3 shows the fetal body segmentation results with the Dice score and ASSD evaluation metrics. Fine tuning with restarts was helpful for both full and partial annotations regimes, increasing the full annotations segmentation Dice score from 0.919 to 0.937 and partial annotations with borders segmentation Dice score from 0.92 to 0.942. Incorporating border information with the selective Dice loss function improved partial annotation setting, increasing the Dice score from 0.936 to 0.942 and decreasing the Dice Standard Deviation (STD) from 0.056 to 0.049. Finally, partial annotations with borders information had slightly better average results to the full annotations regime with a Dice score of 0.937 and 0.942 and ASSD of 3.61 and 3.52 for the full and partial annotations respectively, with a substantially smaller STD: a Dice score STD of 0.063 compared to 0.049 and ASSD STD of 4.04 compared to 3.45 for the full annotations and partial annotations regimes respectively.

**Table 2.** Segmentation results comparison between partial and full annotations for FIESTA body sequence on ID and OOD data. Best results are shown in bold. Unusual behavior for fine-tuning (two step optimization) is indicated with italics.

| Data distribution | Network training | Dice | Hausdorff (mm) | 2D ASSD (mm) |
|---|---|---|---|---|
| **In-Distribution (ID)** | Full | 0.959±0.044 | 34.51±37.26 | 2.15±2.33 |
|  | Full fine-tuned | 0.964±0.040 | 32.98±36.86 | 1.88±2.07 |
|  | Partial | 0.959±0.034 | 34.15±35.96 | 2.21±1.67 |
|  | Partial fine-tuned | 0.965±**0.029** | 31.89±35.82 | 1.90±**1.39** |
| **Out-of-Distribution (OOD)** | Full | 0.836±0.178 | 39.34±29.26 | 7.46±10.61 |
|  | Full fine-tuned | *0.826±0.214* | *39.61±32.66* | *8.86±16.54* |
|  | Partial | 0.875±0.091 | 36.19±21.44 | 5.47±3.92 |
|  | Partial fine-tuned | **0.899±0.067** | **30.37±18.86** | **4.00±2.26** |

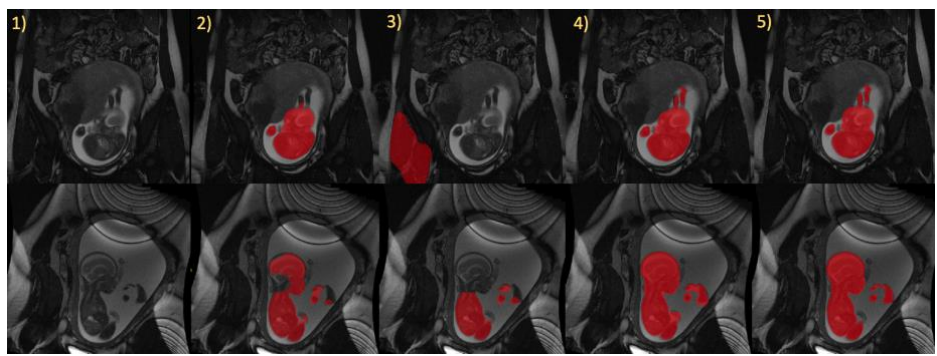

**Fig. 4.** Illustrative fetal body segmentation results for the FIESTA OOD data. Left to right (columns): 1) original slice; 2) Full annotations without fine-tuning; 3) Full annotations with fine-tuning; 4) Partial annotations with fine-tuning; 5) ground truth.

*Study 2: partial annotations for FIESTA sequence for ID and OOD data*

For partial annotations regime, the network was trained on 30 cases and for the full annotations regime the network was trained on 6 cases randomly chosen out of the 30 partially annotated training cases. For both methods, we used the same 6 cases for validation, 68 test cases for ID data and 33 test cases for OOD data.

The OOD data was collected from a different clinical site than the training set and included mostly smaller fetuses (28 out of 33 fetuses had GA between 19-24 weeks compared to GA between 28-39 in the training set). For both partial and full annotations regimes we used Test Time Augmentations (TTA) [17] for the OOD setting to reduce over-segmentation. Because of large resolution differences, we rescaled OOD data to the resolution of 1.56×1.56×3.0 mm$^3$, similar to the resolution of the training set.

In total, eight scenarios were tested, four for ID data and four for OOD data. For both ID and OOD data the following was tested: 1) full annotations without fine tuning. 2) full annotations with fine tuning. 3) partial annotations without fine-tuning. 4) partial annotations with fine-tuning.

Table 2 shows the results. For the ID data, partial annotations results were similar to full annotations with the same annotation effort, but again the STD was much smaller: Dice STD of 0.04 compared to 0.029 and ASSD STD of 2.07 compared to 1.39 for full and partial annotations respectively. For both full and partial annotations regimes the fine tuning slightly improved the segmentation results.

For the OOD data, the differences between segmentation results using full and partial annotations were much larger, with better results for partial annotations regime. Using partial annotations, results improved from a Dice score of 0.836 to 0.899 and from ASSD of 7.46 mm to 4 mm. Unlike in the ID setting, fine-tuning with restarts hurt performance on OOD data in the full annotations regime, potentially indicating an overfitting phenomenon. This was not the case for partial annotations, where again fine tuning with learning rate restarts further improved segmentation results as in the ID setting.

Fig. 4. shows illustrative body segmentation results for the OOD data. Partial annotations showed better performance on these cases compared to full annotations, indicating higher robustness. Also, fine tuning full annotations resulted in decreased performance with a complete failure to the detect the case in the top row, which may indicate an overfitting to the training set.

## 4    Conclusion

We have presented a new method for using partial annotations for large structures. The method consists of algorithm-guided annotation step and a network training step with selective data blocks and a selective loss function. The method demonstrated significantly better robustness under low data regime compared to full annotations.

We also presented a simple two-step optimization scheme for low data regime that combines fine-tuning with learning rate restarts. Experimental results show the effectiveness of the optimization scheme for partial annotations method on both ID and OOD data. For full annotations, the two-step optimization was useful only for ID data but hurt performance on OOD data, indicating potential overfitting.

The selected partial annotations are user-friendly and require only two additional clicks in the beginning and end of the structure of interest, which is negligible compared to the effort required for segmentation delineations. Thus, they can be easily used to construct a dataset with a low annotation cost for initial segmentation network.

## Acknowledgements

This research was supported in part by Kamin Grant 72061 from the Israel Innovation Authority.

## References

1. Reddy UM, Filly RA, Copel JA. Prenatal imaging: ultrasonography and magnetic resonance imaging. Obstetrics and Gynecology 112(1):145-50, 2008.
2. Rutherford M, Jiang S, Allsop J, Perkins L, Srinivasan L, Hayat T, Kumar S, Hajnal J. MR imaging methods for assessing fetal brain development. Developmental Neurobiology 68(6):700-11, 2008.
3. Zhang, T., Matthew, J., Lohezic, M., Davidson, A., Rutherford, M., Rueckert, D et al (2016). "Graph-based whole body segmentation in fetal MR images". Proc. Medical Image Computing and Computer-Assisted Intervention Workshop on Perinatal, Preterm and Paediatric Image Analysis, 2016.
4. Dudovitch G, Link-Sourani D, Ben Sira L, Miller E, Ben Bashat D, Joskowicz L. Deep learning automatic fetal structures segmentation in MRI scans with few annotated datasets. In Proc. Int. Conference on Medical Image Computing and Computer-Assisted Intervention 2020 Oct 4 (pp. 365-374). Springer, Cham.
5. Lo J, Nithiyanantham S, Cardinell J, Young D, Cho S, Kirubarajan A, Wagner MW, Azma R, Miller S, Seed M, Ertl-Wagner B. Cross Attention Squeeze Excitation Network (CASE-Net) for Whole Body Fetal MRI Segmentation. Sensors 21(13):4490, 2021.
6. Tajbakhsh N, Jeyaseelan L, Li Q, Chiang JN, Wu Z, Ding X. Embracing imperfect datasets: A review of deep learning solutions for medical image segmentation. Medical Image Analysis 63(1):101693, 2020.
7. O. Çiçek, A. Abdulkadir, S. S. Lienkamp, T. Brox, and O. Ronneberger, 3D U-net: Learning dense volumetric segmentation from sparse annotation, in Proc. Int. Conf. Med. Image Comput.-Assist. Intervent. Cham, Switzerland: Springer, 2016, pp. 424–432.
8. M. Goetz et al., "DALSA: Domain adaptation for supervised learning from sparsely annotated MR images," IEEE Trans. Med. Imag. 35(1):184–196, 2016.
9. W. Bai et al., "Recurrent neural networks for aortic image sequence segmentation with sparse annotations," in Proc. Int. Conf. Med. Image Comput.-Assist. Intervent. Cham, Switzerland: Springer, 2018, pp. 586–594.
10. L. Lejeune, J. Grossrieder, and R. Sznitman. Iterative multi-path tracking for video and volume segmentation with sparse point supervision. Medical Image Analysis 50:65–81, 2018.
11. Wang S, Nie D, Qu L, Shao Y, Lian J, Wang Q, Shen D. CT male pelvic organ segmentation via hybrid loss network with incomplete annotation. IEEE Trans. Medical Imaging 39(6):2151-62, 2020.
12. Sudre CH, Li W, Vercauteren T, Ourselin S, Jorge Cardoso M. Generalised dice overlap as a deep learning loss function for highly unbalanced segmentations. In Deep learning in medical image analysis and multimodal learning for clinical decision support 2017 Sep 14 (pp. 240-248). Springer, Cham.
13. Smith LN. Cyclical learning rates for training neural networks. In: 2017 IEEE winter conference on applications of computer vision (WACV) 2017 Mar 24 (pp. 464-472). IEEE.
14. Loshchilov I, Hutter F. Sgdr: Stochastic gradient descent with warm restarts. arXiv preprint arXiv:1608.03983. 2016 Aug 13.
15. Kodym, O., Špaňel, M., Herout, A.: Segmentation of Head and Neck Organs at Risk Using CNN with Batch Dice Loss. Lecture Notes in Computer Science 11269 LNCS, 105–114 (2019). https://doi.org/10.1007/978-3-030-12939-2_8
16. Isensee F, Jaeger PF, Kohl SA, Petersen J, Maier-Hein KH. nnU-Net: a self-configuring method for deep learning-based biomedical image segmentation. Nature Methods 18(2):203-11, 2021.
17. Wang G, Li W, Aertsen M, Deprest J, Ourselin S, Vercauteren T. Aleatoric uncertainty estimation with test-time augmentation for medical image segmentation with convolutional neural networks. Neurocomputing 338:34-45, 2019.